\documentclass[conference]{IEEEtran}
\IEEEoverridecommandlockouts
\usepackage{cite}
\usepackage{amsmath,amssymb,amsfonts}
\usepackage{algorithmic}
\usepackage{graphicx}
\usepackage{textcomp}
\usepackage{xcolor}
\usepackage{tabularx}
\usepackage{booktabs}
\usepackage{hyperref}
\usepackage{array}
\usepackage{multirow}
\def\BibTeX{{\rm B\kern-.05em{\sc i\kern-.025em b}\kern-.08em
    T\kern-.1667em\lower.7ex\hbox{E}\kern-.125emX}}
\newcolumntype{Y}{>{\centering\arraybackslash}X}

\begin{document}

\title{Guitar-TECHS: An Electric Guitar Dataset Covering Techniques, Musical Excerpts, Chords and Scales Using a Diverse Array of Hardware}

\author{\IEEEauthorblockN{Hegel Pedroza}
\IEEEauthorblockA{\textit{Dept. of Art \& Humanities} \\
\textit{UAM Lerma}\\
Estado de México, México \\
h.pedroza@correo.ler.uam.mx}
\and
\IEEEauthorblockN{Wallace Abreu}
\IEEEauthorblockA{\textit{Signals, Media \& Telecom. Lab} \\
\textit{UFRJ}\\
Rio de Janeiro, Brasil \\
wallace.abreu@smt.ufrj.br}
\and
\IEEEauthorblockN{Ryan M. Corey}
\IEEEauthorblockA{\textit{Discovery Partners Institute} \\
\textit{UIC}\\
Chicago, United States \\
corey1@uillinois.edu}
\and
\IEEEauthorblockN{Iran R. Roman$^*$}
\IEEEauthorblockA{\textit{School of EECS} \\
\textit{QMUL}\\
London, United Kingdom \\
i.roman@qmul.ac.uk}
}

\maketitle

\begin{abstract}
Guitar-related machine listening research involves tasks like timbre transfer, performance generation, and automatic transcription. However, small datasets often limit model robustness due to insufficient acoustic diversity and musical content. To address these issues, we introduce Guitar-TECHS, a comprehensive dataset featuring a variety of guitar techniques, musical excerpts, chords, and scales. These elements are performed by diverse musicians across various recording settings.
Guitar-TECHS incorporates recordings from two stereo microphones: an egocentric microphone positioned on the performer’s head and an exocentric microphone placed in front of the performer. It also includes direct input recordings and microphoned amplifier outputs, offering a wide spectrum of audio inputs and recording qualities.  All signals and MIDI labels are properly synchronized. Its multi-perspective and multi-modal content makes Guitar-TECHS a valuable resource for advancing data-driven guitar research, and to develop robust guitar listening algorithms. We provide empirical data to demonstrate the dataset’s effectiveness in training robust models for Guitar Tablature Transcription.
\end{abstract}


\section{Introduction}
\noindent Machine learning applications in music listening have significantly benefited music education~\cite{b1,b2,b3}, musicology~\cite{b3,b4,b5,b6}, and performance theory~\cite{b7,b8}. Advances in this area are predominantly data-driven, emphasizing the importance of comprehensive datasets~\cite{b9,b10,b11}.

Addressing the scarcity of electric guitar data---critical for tasks like timbre transfer, music transcription, and performance analysis~\cite{b12,b13,b14}---we introduce Guitar-TECHS. This new dataset features a complete range of guitar notes executed using various techniques, an assortment of chord types, and musical excerpts. It encompasses recordings from multiple guitars performed by diverse professional musicians and utilizes a broad spectrum of recording equipment. Guitar-TECHS is meticulously annotated with labels in MIDI format that detail the onset and duration of each note, offering over five hours of new content. We present empirical evidence of the dataset’s value in guitar tablature transcription (GTT) and propose future work that we hypothesize will greatly benefit from our dataset. We release the dataset and code to reproduce the results presented here with a permissive CC BY 4.0 license\footnote{\noindent \texttt{\url{guitar-techs.github.io}}\\$^{\text{\textbf{\quad*}}}$corresponding author: IR Roman (i.roman@qmul.ac.uk)}. 


\section{Related Work}

\noindent Current guitar datasets offer a range of content, typically curated for specific research tasks. For instance, some datasets provide detailed annotations for musical performance analysis, like guitar fingering~\cite{b15}, or for detecting solos in rock performances~\cite{b16}. Among these, GuitarSet~\cite{b17} stands out for its authentic guitar performances and precise note annotations, contrasting with datasets that simulate guitar performances programmatically~\cite{b18,b19,b20} or focus on emulating or capturing guitar hardware~\cite{b21,b22}. However, GuitarSet, with its three-hour content primarily of acoustic guitar, is considered small by today's machine learning dataset size standards.

\begin{figure}[t] 
    \centering
    \includegraphics[width=0.95\columnwidth]{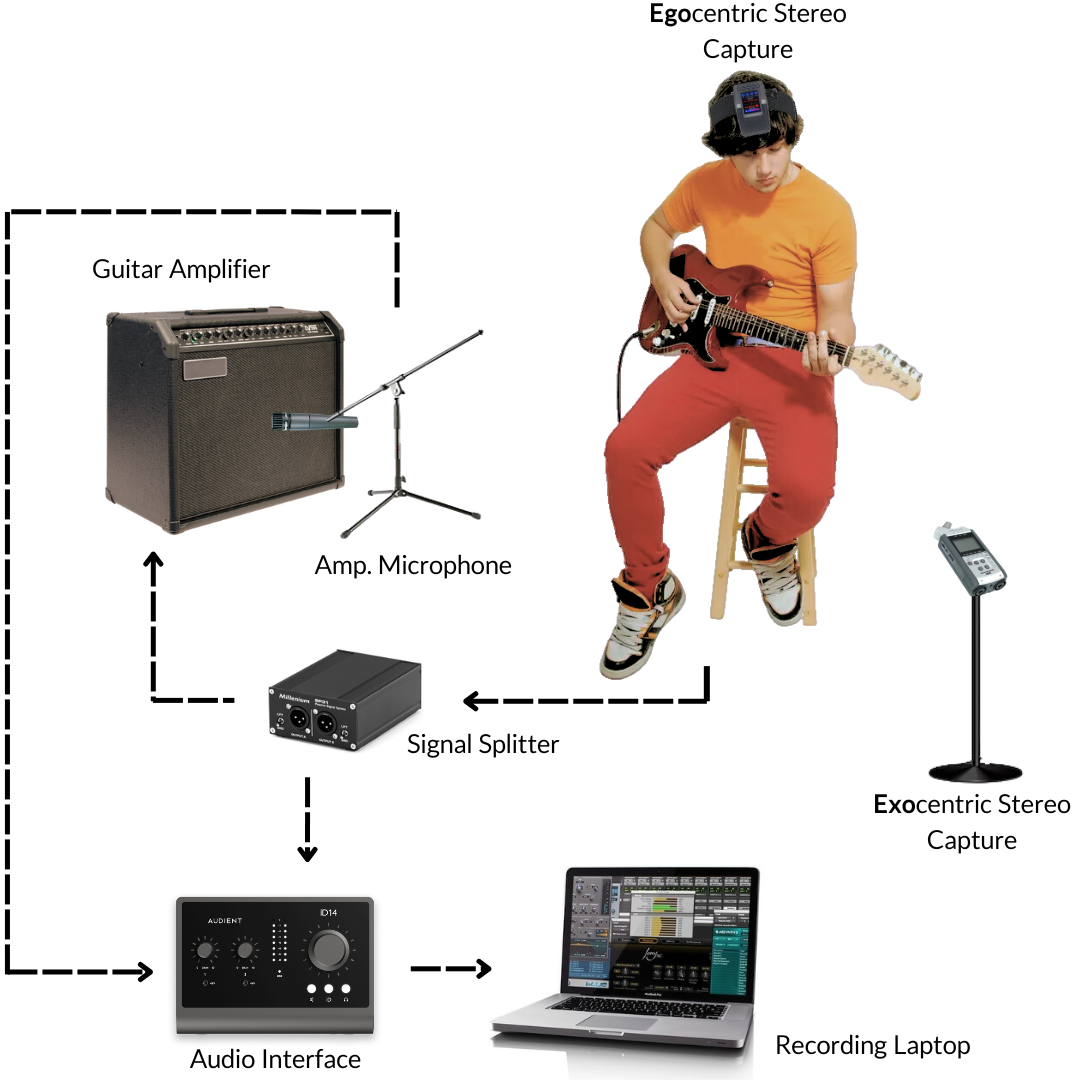} 
    \caption{Depiction of the recording setup used. Note that the egocentric and exocentric microphones---respectively functioning as the listening perspective of the player and a hypothetical audience---captured stereo signals.}
    \label{fig:side_by_side_images}
\end{figure}

There is also increasing interest in datasets that document the same phenomena from multiple perspectives, such as the player’s and the audience’s~\cite{b23,b24,b25}, which are valuable for multiview learning and augmented and virtual reality applications~\cite{b30,b31,b32,b33,b34,b35}. Yet, a specialized guitar dataset catering to multi-perspective listening is still lacking. Our proposed Guitar-TECHS dataset fills this gap by incorporating four distinct audio collection setups: direct input to a computer, a miked amplifier, an egocentric microphone, and an exocentric microphone (see Fig. \ref{fig:side_by_side_images}).

\section{Data Collection Process}
\label{sec:pagestyle}

\noindent The dataset consists of guitar content by three professional players (\texttt{Player 01}, \texttt{Player 02}, \& \texttt{Player 03}), each recording on separate dates. \texttt{Player 01} and \texttt{Player 02} contributed recordings of various scales, chords, and tone-playing techniques, while \texttt{Player 03} focused solely on music excerpts. The recordings took place in different rooms within a house, utilizing distinct guitars and amplifiers. We also varied the amplifier microphones and computer interfaces to diversify the audio inputs. As a result, each player’s recordings reflect unique guitar tonal qualities and are captured under varying acoustic conditions in household environments.

\subsection{Recording setup}

\noindent All three guitars feature 22 frets and six standard-tuned strings (E-A-D-G-B-e). Guitars were equipped with a Fishman Triple Play Connect multi-track MIDI pickup, allowing for the independent and automated capture of MIDI notes from each string.
The audio output from each guitar was routed through a signal splitter, allowing us to send the signal to an audio interface for direct digital collection (i.e. ``direct input'') and an amplifier that was recorded using a microphone connected to the same audio interface.
Additionally, audio was captured from two perspectives: a player perspective (or egocentric) using a head-mounted stereo microphone, and a listener (or exo-centric) perspective also using a stereo microphone but positioned 5 feet directly in front of the performer. This distance was used because it reflects a typical teacher-student guitar lesson scenario, which is an application that we envision our dataset could enable. 
Fig. \ref{fig:side_by_side_images} depicts the recording setup. 

\subsection{Specific hardware settings per guitar player}

\noindent Table~\ref{table:sample} gives an overview of the hardware used by each player. Here we give a more detailed description of recording room conditions and detailed hardware settings.

\texttt{Player 01}: Performed in a bedroom using an Ibanez Performer PF-300 Les Paul type guitar, fitted with 0.011-0.050 inch flat-wound strings. Its tone and volume controls were set to maximum, and the bridge pickup was used throughout.
The signal was amplified through an Orange CR60 combo amp and captured using a Shure SM57 dynamic cardioid mic positioned on-axis to the speaker cone. 

\texttt{Player 02}: Performed in a living room office using an EVH Wolfgang Exotic Super Stratocaster type guitar, fitted with 0.009-0.42 inch nickel-wound strings. Its volume and tone controls were set to maximum, and only the neck pickup was used.
The signal was amplified through a Yamaha YB15 bass amp and captured using an Audio-Technica AT2020 condenser cardioid mic. 

\texttt{Player 03}: Performed in a dining room using a Sire T7 Telecaster type guitar, fitted with 0.010-0.046 inch nickel-wound strings. Its tone and volume controls were set to maximum, and the middle pickup selector position was used, enabling simultaneous use of both the bridge and neck pickups.
The signal was amplified through a CR-12 combo amp and captured using a Behringer ECM8000 omnidirectional measurement mic. 

All amplifiers were set to a flat EQ, with volume adjusted by each player to reflect their playing and listening preference.

To capture a wide range of audio tones and recording techniques, two different audio interfaces – an Audient iD14 and a Behringer UMC202HD – were used throughout the recording process. Additionally, all recordings were captured using two digital audio workstations (DAWs), Reaper and Pro Tools. The recordings were conducted on a Lenovo Y720 computer. This diverse recording setup aimed to simulate conditions often encountered in home recording setups, rather than professional studios.

\begin{table}[t]
\centering
\caption{Overview of Hardware Used by Each Player}
\label{table:sample}
\begin{tabular}{|l|c|c|c|}
\hline
\textbf{Hardware} & \textbf{Player 01} & \textbf{Player 02} & \textbf{Player 03} \\
\hline
\hline
Guitar & Ibanez PF300 & EVH Wolfang & Sire T7 \\
\hline
Strings Guge & 0.011 - 0.050 & 0.009 - 0.042 & 0.010 - 0.046 \\
\hline
Pickup & Bridge & Neck & Bridge-Neck \\
\hline
Amplifier & Orange CR-60 & Yamaha B-15 & Orange CR-12 \\
\hline
Amp Mic & SM57 & AT-2020 & ECM8000 \\
\hline
Audio Interface & ID 14 & UMC202HD &  UMC202HD \\
\hline
\end{tabular}
\end{table}

\section{Data Content}
\label{sec:typestyle}

\noindent The dataset is organized into four categories: Techniques, Musical Excerpts, Chords and Scales. The following paragraphs better describe the content of each. 

\textbf{Techniques:} Includes recordings of six techniques to play single guitar notes: Alternate Picking, Palm Mute, Vibrato, Harmonics, Pinch Harmonics, and Bendings. 
For Single Notes, Palm Mute, Vibrato, and Pinch Harmonics, all possible fretted notes and open strings (138 total) were recorded, with each tone lasting a total of 4 seconds. 
For Bendings and Harmonics, performers were given creative freedom to execute the techniques in a manner that best captured their desired expression, resulting in varying durations and numbers of events. Note that we include ground-truth MIDI annotations for all tones.

\textbf{Chords:} Includes triads (major, minor, augmented, and diminished) and seventh chords (major 7, minor 7, dominant 7, and minor 7 flat 5). Each was recorded in root position and with inversions across different string sets. 
Triads were performed using close voicings across strings set 1, 2, 3, and 4, while seventh chords were recorded with drop 2 voicings on strings sets 1 and 2, and drop 3 voicings on strings 2, 3, 4 and 6. 
Each chord was performed using alternate strumming, starting from its lowest voicing and ascending chromatically up to the 12th fret, with a duration of 4 seconds per chord.

\textbf{Scales:} Captures the performance of individual notes on the guitar, but with a sequential pattern that is not a simple chromatic sequence. Encompasses all major scales in the twelve keys, performed in ascending and descending box patterns using alternate picking. Each scale box pattern has between 15 and 17 different notes, with each note held for 0.5 seconds.

\textbf{Original music performances:} Encompasses complete and original musical passages in the form of guitar solos. These provide a broader and realistic context for potential analysis of performance elements such as phrasing, rhythm, and articulation. A total of 12 musical excerpts showcasing a diverse range of tempos, techniques, and musical elements, including the use of a guitar capo and finger-plucking techniques.

\subsection{Data format and specifications}

\noindent All data was simultaneously recorded using an egocentric and an exo-centric microphone. This in addition to the audio captured directly from the guitar and amplifier. The corresponding MIDI annotations provide precise labeling for each musical event. The format of this data is therefore:
\begin{itemize}
    \item Audio: wav format with a sampling rate of 48,000 Hz and a bit depth of 32-bit floating point.
    \item MIDI: Multitrack format consisting of six tracks with a resolution of 9600 ticks per quarter note.
\end{itemize}
All files were systematically named and indexed according to their categories. Table~\ref{tab:tabII} provides a breakdown of the dataset's content and total durations when accounting for data collected across players. 

\begin{table}[t]
\centering
\caption{Summary of Dataset Content}
\small
\begin{tabularx}{\columnwidth}{|l|Y|Y|Y|c|}
\hline
& \multicolumn{3}{c|}{\textbf{Elements per Player}} & \\
\cline{2-4}
& \textbf{Player 01} & \textbf{Player 02} & \textbf{Player 03} & \textbf{Total Dur.}\\
\hline
\hline
\textbf{Techniques:} & & & &\\

- Single Notes & 138 & 138 & 0 & 00:18:24 \\
- Palm Mute & 138 & 138 & 0 & 00:18:24 \\
- Vibrato & 132 & 132 & 0 & 00:17:36 \\
- Pinch Harmonics & 132 & 132 & 0 & 00:17:36 \\
- Harmonics & 30 & 30 & 0 & 00:04:00 \\
- Bendings & 30 & 30 & 0 & 00:04:00 \\

\hline
\textbf{Musical Excerpts:} & 0 & 0 & 12 & 00:08:02 \\
\hline
\textbf{Chords:} & & & & \\
- Three-Note Chords & 624 & 624 & 0 & 01:23:12 \\
- Four-Note Chords & 576 & 576 & 0 & 01:16:48 \\

\hline
\textbf{Scales:} & 12 & 12 & 0 & 01:04:00 \\
\hline
\textbf{Grand Total:} & 1860 & 1860 & 12 & 05:12:02 \\
\hline
\end{tabularx}
\label{tab:tabII}
\end{table}


\begin{table*}[ht!]
\centering
\caption{Results comparing the original TabCNN model against one trained with the inclusion of Guitar-TECHS in the training data. The average and std across validation folds are shown. Boldface values indicate the best average per metric.}

\label{tab:metrics}
\begin{tabularx}{\textwidth}{@{}l *{7}{>{\centering\arraybackslash}X}@{}}
\toprule
 & \multicolumn{3}{c}{\textbf{Multi-pitch estimation}} & \multicolumn{3}{c} {\textbf{Tablature estimation}} & \\
\cmidrule(lr){2-4} \cmidrule(lr){5-7}
& \textbf{F\textsubscript{1}} & \textbf{P} & \textbf{R} & \textbf{F\textsubscript{1}} & \textbf{P} & \textbf{R} & \textbf{TDR} \\ \midrule
TabCNN\cite{b36} & 0.826$\pm$0.025 & 0.900$\pm$0.016 & 0.764$\pm$0.043 & \textbf{0.748}$\pm$0.047 & \textbf{0.809}$\pm$0.029 & 0.696$\pm$0.061 & 0.899$\pm$0.033 \\
\hline
+ Guitar-TECHS & \textbf{0.828}$\pm$0.023 & 0.\textbf{909}$\pm$0.021 & \textbf{0.767}$\pm$0.044 & 0.747$\pm$0.031 & \textbf{0.809}$\pm$0.018 & \textbf{0.699}$\pm$0.048 & \textbf{0.905}$\pm$0.015 \\
\bottomrule
\end{tabularx}
\label{tab:III}
\end{table*}

\begin{table}[ht!]
\centering
\caption{Tablature Disambiguation Rate on EGSet12. Comparison of the base model and the model trained with Guitar-TECHS.}

\begin{tabular}{ccc}
 & \textbf{TDR}\\
\hline
TabCNN~\cite{b36} & 0.695$\pm$0.075\\
\hline
+ Guitar-TECHS & \textbf{0.776}$\pm$0.106 \\
\hline
\end{tabular}
\label{tab:IV}
\end{table}

\section{Experiments Using Guitar-TECHS}

\noindent We now demonstrate the value of our dataset for empirical guitar research. 
We used Guitar-TECHS data to increase the training splits of GuitarSet~\cite{b17} and train TabCNN\cite{b36}---a model for automatic Guitar Tablature Transcription (GTT) with a convolutional neural network (CNN) architecture. We trained two TabCNN models, one trained using Guitar-TECHS and one without, using k-fold cross-validation on GuitarSet. After training with GuitarSet, we also evaluate models on the recently-published EGSet12 dataset~\cite{b22}, which was specifically created to evaluate and compare GTT model performance in a new data domain/distribution. 

Our assessment is therefore two-fold. First, we determine whether incorporating Guitar-TECHS into the original TabCNN training process results in improved model performance compared to using GuitarSet alone.
Second, we evaluate model generalization for GTT as a function of including Guitar-TECHS data. 

\subsection{Training TabCNN models with GuitarSet \& Guitar-TECHS}

\textbf{Training methodology.} We employed the AMT Tools version of TabCNN~\cite{b37} and followed the six-fold cross-validation scheme from GuitarSet to train two distinct models. One model was trained using only GuitarSet, while the other incorporated Guitar-TECHS into the training data splits, ensuring GuitarSet was used solely for cross-validation to maintain a fair comparison~\cite{b17}. All other training parameters, such as model architecture, optimizer, learning rate, batch size, and validation data, remained unchanged from the AMT implementation of TabCNN~\cite{b36}.

\textbf{Metrics.} We used a dual-focus approach, analyzing both multi-pitch accuracy and tablature precision, as outlined by Wiggins \& Kim~\cite{b36}. Metrics included F1 score, precision, and recall for both types of metrics, with the addition of tablature disambiguation rate (TDR) to evaluate correct fret and string assignments.

\textbf{Results.} Training with Guitar-TECHS resulted in modest improvements across multi-pitch metrics, evidenced by better F1 scores, precision, and recall. Tablature metrics showed stable F1 scores and precision, with a slight increase in recall. The tablature disambiguation rate also saw a small improvement. These findings, summarized in Table \ref{tab:metrics}, indicate that Guitar-TECHS can enhance the training of GTT models.


\subsection{Assessment of Model Robustness with EGSet12}

\textbf{Methodology}. We tested both models on the EGSet12 dataset~\cite{b22,b32}, which differs significantly from GuitarSet and Guitar-TECHS in terms of recording hardware and acoustic conditions. Our hypothesis was that the model trained with Guitar-TECHS, featuring diverse acoustic environments and hardware, would more accurately identify pitch corresponding to specific guitar frets. TDR is the key metric for this assessment, as it measures a model’s accuracy in pinpointing pitch on the correct string and fret.

\textbf{Results.} The model that incorporated Guitar-TECHS into its training showed a notable improvement in TDR, exceeding the baseline model by more than 8 points as detailed in Table~\ref{tab:IV}. These findings confirm that Guitar-TECHS significantly enhances the model’s ability to accurately determine pitch and tablature positions, supporting our initial hypothesis.

\section{Conclusions}
\label{sec:page}

\noindent We have introduced Guitar-TECHS, a new dataset that integrates multi-perspective audio recordings with MIDI annotations of guitar techniques, musical excerpts, chords and scales. To evaluate the dataset's effectiveness, experiments were conducted on the specific task of guitar tablature transcription (GTT).
Results showed that our dataset can enhance model performance, both during in-domain cross-validated training and out-of-domain evaluation, highlighting the value of our dataset in enhancing the model's accuracy and generalization.


\section{Future Work}

\textbf{Guitar listening research}. Guitar-TECHS offers extensive possibilities for advancing guitar-related AI technologies. A direct application could be improving guitar tone and reducing reverberation through the diverse sound qualities captured in our dataset, which could also support the generation of synthetic data to enhance GTT model robustness~\cite{b22}.
Utilizing the dataset’s ego-exo audio captures, “cross-listening” techniques, similar to cross-view learning~\cite{b23}, could be applied to establish connections within the different listening perspectives of the same musical events that Guitar-TECHS provides. This approach would foster the development of more accurate models for identifying different guitar players or techniques in live performances and differentiating roles such as soloist versus accompanist.

\textbf{Augmented and virtual reality}. In augmented reality (AR), multiple audio perspectives are vital for creating immersive depictions of guitar playing~\cite{b34}. Our dataset, with its rich and varied audio captures, is ideally suited to enhance the realism and interactivity of AR experiences. Similarly, in virtual reality (VR), the dataset’s high-fidelity audio inputs can significantly enrich virtual guitar performance experiences, offering more realistic and engaging sounds in virtual environments~\cite{b35}.

\textbf{Model interpretability}. The dataset could also be used to enhance interpretability of music information retrieval (MIR) systems. Using its detailed annotations and diverse audio inputs, researchers could develop methods to analyze how MIR systems process varying guitar techniques and tones. This approach would facilitate a clearer understanding of the decision-making processes in automated music transcription and analysis, promoting more reliable grounding of models.

\textbf{Future enhancements}. Moving forward, we plan to broaden the dataset to cover more musical genres and playing styles to increase its applicability. Additionally, we aim to integrate motion capture data, video recordings, and human physiological measurements. This expansion will provide a deeper understanding of the cognitive and physical aspects of guitar performance and the relationship between a performer’s physical actions and physiological processes. Further refinement of MIDI annotations through subjective assessments and human annotators will also be pursued to improve the dataset’s precision and utility.

\section{Acknowledgments}
\label{sec:illust}

\noindent This research was supported by the IEEE Mentoring Experiences for Underrepresented Young Researchers (ME-UYR) program and Mexico's National Scholarship for Graduate Studies from the National Council of Humanities, Sciences, and Technologies (CONAHCYT). The authors extend their gratitude to these funding sources and collaborators, with special thanks to Dr. Hugo Solís García for his valuable feedback and for providing access to computational resources through the CONAHCYT “Ciencia de Frontera, Paradigmas y Controversias de la Ciencia 320551” program. Additionally, we sincerely thank guitarists David Pang and Rodolfo Samperio, who, alongside the first author, contributed to the recording of all the data presented.

\end{document}